\documentclass[epj,final]{svjour}
\usepackage{graphics}
\usepackage{psfig}
\usepackage{amssymb}

\begin{document}
\title{Molecular Dynamics of Comminution in Ball Mills}
\author{Volkhard Buchholtz\inst{1}, Jan A. Freund\inst{2}, Thorsten
  P\"oschel\inst{2,3}} \institute{
  \inst{1} Logos Verlag Berlin, Michaelkirchstr. 13, D-10179 Berlin, 
Germany. \email{v.buchholtz@logos-verlag.de}\\
  \inst{2} Humboldt-Universit\"at, Institut f\"ur Physik,
  Invalidenstr. 110,
  D-10115 Berlin, Germany. \email{freund@physik.hu-berlin.de}\\ 
http://summa.physik.hu-berlin.de/$\sim$kies/\\
  \inst{3} ICA1, Universit\"at Stuttgart, Pfaffenwaldring 27, D-70569
  Stuttgart, Germany. \email{thorsten@physik.hu-berlin.de}}
\date{January 19, 2000}
\abstract{
  We investigate autogenous 
fragmentation of dry granular material in
  rotating cylinders using two-dimensional molecular dynamics. By
  evaluation of spatial force distributions achieved numerically for
  various rotation velocities we argue that comminution occurs mainly
  due to the existence of force chains. A statistical analysis of
  theses force chains explains the spatial distribution of comminution
  efficiency in ball mills as measured experimentally by
  Rothkegel~\cite{Rothkegel:1992} and Rolf~\cite{Rolf}. For animated sequences of
  our simulations see
  http://summa.physik.hu-berlin.de/$\sim$kies/Research/RotatingCylinder/rotatingcylinder.html
}
\maketitle
\section{Introduction}
In the past years molecular dynamics simulations of granular systems
-- such as sand, fertilizer, grain and others -- have been developed
towards a reliable tool for an investigation of granular systems.
Unlike a few years ago, when physicists started to investigate
granular assemblies consisting of just a few hundred spherical
granular particles in two dimensions,
e.g.~\cite{HW,CS,Herrmann,Taguchi,Walton,WaltonBraun,BuchholtzPoeschel:1997},
and less than 100 particles in three dimensions, e.g.~\cite{GHPS},
today, owing to the rapid evolution of computer facilities, we are
able to simulate complex systems of many thousands of particles in two
and three dimensions even accounting for non-spherical grain features.

Due to this progress molecular dynamics of granular systems by now can
be applied to the simulation of technologically important processes
striving for a better understanding of relevant details of the process
and, henceforth, for an optimization of industrial technologies
(e.g.~\cite{Kohring:1993,SEP,SEP1,Kohring,YokoyamaTamuraJimbo:1994,YokoyamaTamuraUsuiJimbo:1996,SongfackAgrawalaMishraRajamani:1993}).
Molecular simulation techniques, therefore, might develop to a
prospective engineering tool partly replacing costly laboratory
experiments by ``computer experiments''.

In the present paper we apply the method of molecular dynamics to the
simulation of autogenous 
dry comminution in a tumbling ball mill \cite{ballmill}.
To this end we devised a novel algorithm which accounts for the
fragmentation of particles. First we will demonstrate that
experimentally known results, in particular those by
Rothkegel and Rolf~\cite{Rothkegel:1992,Rolf}, can be reproduced up to a
good degree of accuracy. We will show that the efficiency of
comminution in ball mills is mainly determined by the presence of
force chains. Our result will hint at how to improve the efficiency of
a milling machinery widely applied in industry.
\section{The Simulation Model}
\subsection{Molecular Dynamics}
For the simulation we assumed pairwise interaction forces and applied
the model by Cundall and Strack~\cite{CS} and Haff and
Werner\cite{HW}: A pair of two-dimensional spherical particles $i$ and
$j$ interacts through the force
\begin{equation}
  \label{force}
  \vec{F}_{ij} = F_{ij}^N\, \vec{n} + F_{ij}^T\, \vec{t} ,
  \label{force1}
\end{equation}
if their distance is smaller than the sum of their radii
$R_i+R_j-\left| \vec{r}_i-\vec{r}_j\right| = \xi_{ij} >0$. The unit
vectors in normal and tangential direction are defined as
\begin{eqnarray}
  \vec{n} &=& \vec{r}_{ij}/\left|\vec{r}_{ij}\right|\\[1ex]
  \vec{t} &=& \left( {0 ~ -1}\atop{1 ~~~~ 0}  \right) \cdot 
\vec{r}_{ij}/\left|\vec{r}_{ij}\right|\,,
\end{eqnarray}
with $\vec{r}_{ij}\equiv\vec{r}_i-\vec{r}_j$.
The forces in normal and tangential direction are respectively given
by
\begin{eqnarray}
  F_{ij}^N &=& Y \xi_{ij} - m_{ij}^{{\rm eff}}\, \gamma_N\, 
  \dot{\vec{r}}_{ij} \cdot\vec{n} \label{fnormal}\\
  F_{ij}^T &=&  \mbox{sign} (v_{ij}^{\rm rel})\,
  \min\left( m_{ij}^{{\rm eff}} \gamma_T \left|v_{ij}^{\rm rel}\right|
    , \mu \left| F_{ij}^{N}\right| \right) \label{ftang} 
\end{eqnarray}
with
\begin{eqnarray}
  v_{ij}^{\rm rel} &=& \dot{\vec{r}}_{ij} \cdot
  \vec{t}+ R_i \, \Omega_i + R_j \, \Omega_j 
  \label{surfvelocity}\\ 
  m_{ij}^{{\rm eff}} &=& \frac{m_i \, m_j}{m_i + m_j} ~.
  \label{meff}
\end{eqnarray}

Here $R_i$, $m_i$, $\vec{r}_i$, $\dot{\vec{r}}_i$ and $\Omega_i$ are
respectively the radius, the mass, the position, the velocity and the
rotation velocity of the $i$th particle.  $Y=8\cdot 10^6$\,g\,s$^{-2}$ is
the elastic constant and $\gamma_N=800$\,s$^{-1}$, $\gamma_T=3000$\,s$^{-1}$
are the parameters of damping in normal and tangential direction. The
Coulomb friction constant was assumed to be $\mu=0.5$. These empirical
values which we used throughout all simulations have proven to render
realistic behaviour for a typical granular material.

The normal force (\ref{fnormal}) is composed of an elastic repulsive
part and a dissipative part which acts against the direction of
motion. For a collision of two-dimensional spheres, i.e.~disks, the
Hertz contact law $F_{ij}^N\sim \xi_{ij}^{3/2}$
\cite{Hertz:1882,Landau:1965} reduces to \cite{Engel}
\begin{equation}
  \label{2dhertz}
  \xi_{ij} \sim F_{ij}^N \left( {2\over 3} +
    \ln {4 (R_i+R_j)E_r\over F_{ij}^N} \right)~,
\end{equation}
where $E_r$ is the reduced elastic module, i.e.~a material constant.
Equation (\ref{2dhertz}) provides a relation
$\xi=\xi\left(F^N\right)$, however, in the molecular dynamics
simulations we need the inverse $F^N=F^N(\xi)$. To calculate the
forces due to Eq. (\ref{2dhertz}) one would have to invert
(\ref{2dhertz}) numerically for each particle contact in each time
step, i.e., to solve a transcendental equation. Alternatively one
could tabulate the function $F^N(\xi)$ but due to large force
gradients, in particular in the instant of contact, we have to
simulate with double precision accuracy. Therefore, the table would be
extremely large.  Apart from the small logarithmic term $-F_{ij}^N\ln
F_{ij}^N$ in Eq. (\ref{2dhertz}) one finds that the force is
proportional to the compression $\xi_{ij}$.  Therefore, in our
simulations we used the linear law (\ref{fnormal}) which is a good
approximation in the force interval of interest.

Equation (\ref{surfvelocity}) describes the relative velocity of the
particle surfaces at the point of contact which results from both the
tangential part of the relative particle velocity and the velocity of
particle spin.

The Coulomb friction law is taken into account in Eq.~(\ref{ftang}):
If the tangential force of two colliding particles exceeds $\mu$ times
the normal force the particles slide upon each other feeling constant
friction. In this way the Coulomb law formulates a maximum transferable
shear force.

As numerical integration scheme we used the Gear predictor-corrector
method of sixth order (e.g.~\cite{AllenTildesley}).

Let us remark that there exist various models with diverse
descriptions of the interaction forces, e.g.~\cite{WolfKraefte}. In
three-dimensional calculations it is crucial that the dissipative term
in (\ref{fnormal}) is proportional to
$\sqrt{\xi}\dot{\xi}$~\cite{BSHP} (in viscoelastic approximation). In
three dimensions the used term $\sim\dot{\xi}$ in Eq.~(\ref{fnormal})
yields wrong results \cite{WolfKraefte,rospap}.

Another class of molecular dynamics are event-driven simulations,
e.g.~\cite{Rapaport} which require much less numerical effort than the
method described above (force-driven Molecular Dynamics). In
event-driven simulations one does not integrate the equations of
motion explicitly but rather expresses the loss of mechanical energy
in normal and tangential direction due to a collision via coefficients
of restitution $\epsilon^N$ and $\epsilon^T$ both of which can be
determined from the material properties of the particles
\cite{BSHP,SchwagerTP2}. Besides some conceptional difficulties which
can be overcome by applying numerical tricks \cite{luding} these
algorithms are only applicable in case the duration of collisions is
negligible in comparison with the mean free flight time, i.e.~the time
a particle moves without interaction. In this limit collisions can be
assumed as instantaneous events. Moreover, this implies that
three-particle interactions are very rare events. However, in the
system under study, namely the ball mill, this premise does not hold
because the majority of particles permanently is in close contact with
neighbouring particles.
\subsection{Fragmentation Probability of Grains}
\label{sec:2.2}
It can be observed in experiments that different particles of
identical size and material vary with respect to their tensile
strengths. That means when compressing particles of a sample by
applying a fixed force only a fraction of the sample actually will
break. For a sufficiently large sample this fraction can be rephrased
in terms of a fracture probability.

The reason for this different behavior can be explained by the
existence of flaws which provide sites for stress concentration and
the initiation of a crack which, subsequently, propagates through the
material. These flaws are distributed throughout the whole volume,
however, several investigations have revealed that surface flaws
activated by high tensile stress play the dominant role for the
initiation of a crack~\cite{Weichert:1992}. Hence, the fracture
probability is intimately related to the statistical distribution of
surface flaws and the resulting stress distribution over the
particle's surface.

Starting from some plausible assumptions concerning the flaw
distribution and the behaviour of the material the fracture
probability $P$ of a single particle can be derived employing
statistical reasoning. A particle of radius $R$, subjected to a force
which stores the specific elastic energy $W_m=W/m$, will break with
probability \cite{Weichert:1988}
\begin{equation}
  \label{probbreak}
  P(R,W_m) \sim 1-\exp\left(-c R^2W_m^z\right)~.
\end{equation}
Here, $c$ and $z$ are material constants. The original derivation
(considering rods of a brittle material) goes back to Weibull
\cite{Weibull:1939}. This simple law has proven in practice by fitting
various experimental data. The constants $c$ and $z$ can be extracted
from measurements when plotting $[\ln\ln (1-P)^{-1}]$ vs.~$[\ln W_m]$.
Typical values of $z$ are found to lie in
the range $1.5,\ldots,2.5$.

Notice that in agreement with experimental results the term $R^2$ in
the exponent (for fixed $W_m$) in general predicts an anticorrelation
between particle size and resistance to breakage. Clearly, this fact
is rooted in the diminished probability to find a sufficiently large
flaw on a smaller surface.

For spherical particles (or, more generally, for particles having
convex surface), assuming linear elastic behaviour, Hertz-theory
\cite{Hertz:1882,Landau:1965} can be employed to derive a relation
between the elastic energy $W$ stored in a sphere and the exerted
repulsive contact force $F$
\begin{equation}
  \label{Hertz}
  W = {2\over 5}hF={2\over 5}\left({D^2F^5\over R}\right)^{1\over 3}~,
\end{equation}
where $h$ quantifies the particle's deformation and $D$ is another
material constant (expressed by the Poisson ratio and constant of
elasticity) \cite{Hertz:1882,Landau:1965}. It should be kept in mind
that formula (\ref{Hertz}) remains valid in the linear elastic regime
only.

\begin{figure}[htbp]
  \centerline{\psfig{figure=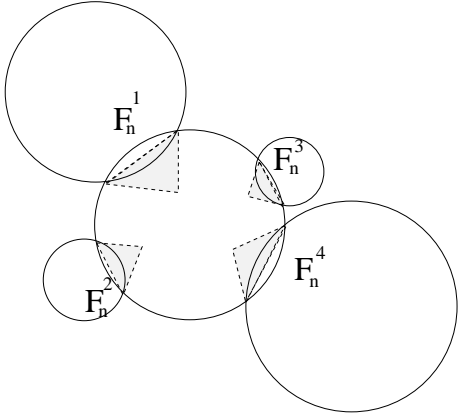,width=6.0cm}}
  \caption{A typical situation of one particle being compressed by
    four impacting particles. The forces are considered as independent
    stress sources. Therefore, in simulations we evaluated only the
    maximum of all exerted forces to compute the fracture probability
    $P(R,W_m)$.}
  \label{fig:typical}
\end{figure}

Applied to our simulation of ball mills we make use of these laws in
the following way: Consider the situation sketched in
Fig.~\ref{fig:typical}: A particle placed in the centre is compressed
by four impacting particles.

The highest compressive stress occurs around the contact point; a fact
well known to engineers \cite{BernotatSchoenert:1988} and also found
in numerical simulations \cite{KunHerrmann:1996,KunHerrmann:1996a}.
Hence, the fracture probability of the centre particle is linked to
the probability of finding a flaw of sufficient size in the very
vicinity of any of the four contact regions. Thus, the action of
different impacting particles can be regarded as independent stress
sources. That means we consider the elastic energy stored by each
impacting particle separately, calculate the related fracture
probability according to formulae (\ref{probbreak}) and (\ref{Hertz})
as a function of the particle's size and the maximum of all acting
normal forces $F^N$
\begin{equation}
  \label{probbreakfinal}
  P(R,F^N) \sim 1-\exp\left(-\tilde{c} R^{\left(2-z{10\over 3}\right)} 
    \left(F^N\right)^{5z\over 3}\right)~,
\end{equation}
and actually decide the question of breakage by subsequently drawing a
random number for each of the contact sites.

The constant $\tilde{c}$ is related to the empirical constant $c$ in
Eq.~(\ref{probbreak}); the precise connection is elaborated by
inserting Eq.~(\ref{Hertz}) into (\ref{probbreak}) and expressing $m$
by the particle radius $m\propto R^3$.

Up to this point we discussed the fracture probability for
three-dimensional spheres. In our simulations we modelled
two-dimensional spheres, i.e.~circular discs of unique thickness $L$.
The fracture probability corresponding to Eq.~(\ref{probbreakfinal})
can be obtained in analogy to (\ref{probbreak}) yielding
\begin{equation}
  \label{probbreak2}
  P(R,W_m) \sim 1-\exp\left(-c R L  W_m^z\right)~.
\end{equation}
Note that the surface of a sphere $\propto R^2$ has to be replaced by
the surface of a cylinder $\propto R L$ (excluding top and bottom
wall). Correspondingly Eq.~(\ref{Hertz}) turns into $h = Y F^N$ and,
hence, the elastically stored energy for spheres has to be replaced by
\begin{equation}
  \label{Hertz2}
  W = {1\over 2}h F^N = {(F^N)^2\over 2 Y}\sim (F^N)^2~.
\end{equation}
Taking into account that the mass of a cylinder is proportional 
$\propto R^2$, one obtains
\begin{equation}
  \label{probab}
  P(R,F^N) \sim 1-\exp\left(-\tilde{c} R 
    \left[{(F^N)^2 \over R^2}\right]^z\right)~.  
\end{equation}
In section \ref{sec:2.3} we will motivate the choice $z=2$. With this
specification one finally arrives at the following expression for
the fracture probability
\begin{equation}
  \label{probbreakfinale}
  P(R,F^N) \sim 1-\exp\left(-\tilde{c} R^{-3} \left(F^N\right)^4\right)~.
\end{equation}
\section{Fragment size distribution}
\label{sec:2.3}
Once a flaw has been activated it forms an initial crack which rapidly
propagates through the particle. Typical crack velocities in the range
of $1500 m/s$ have been measured~\cite{Weichert:1992}. The formation
of a variety of differently sized fragments occurs by virtue of
branching cascades. Branching is governed by a balance between
the energy release rate $G$ and the so-called crack resistivity
$B$~\cite{Weichert:1992} the latter accounting for the creation of new
surfaces. Because of the extremely short duration of the rupture
process one can safely neglect external energetic contributions to
this energy balance. The energy release rate $G$ depends on the crack
velocity and the crack length whereas the crack resistivity $B$
depends on the crack velocity and on the number of branches. Due to a
maximal crack velocity, i.e.~the speed of sound, the balance between
both terms requires the formation of new cracks. From these
considerations it can be shown \cite{Weichert:1992} that the
fragment size distribution (cumulative mass distribution) $Q$ is a
function of the product $R_f\, W_m$, where $R_f$ denotes the
fragment size, i.e.~$Q=Q(R_f\, W_m)$.

Experimental evidence \cite{Weichert:1988} for a scaling law $Q\sim
R_f^{\beta}$ is in agreement with the well-known empirical Schuhmann
law~\cite{Schuhmann:1940}
\begin{equation}
  \label{schumann}
  Q\sim\left({R_f\over k}\right)^{\beta}~,
\end{equation}
which itself can be regarded as an approximation of the Rosin-Rammler
law~\cite{RosinRammler:1933}
\begin{equation}
  \label{rosin_rammler}
  Q\sim 1-\exp\left[-\left({R_f\over k}\right)^{\beta}\right]\,.
\end{equation}
The variable $k$ has the dimension of a length and, thus, can only be
identified with the size of the original particle.

A rigorous derivation of Eq.~(\ref{rosin_rammler}) -- compatible with
an exponent $\beta =1$ -- starting from rather mild assumption and
applying Poisson statistics was performed by Gilvary
\cite{Gilvary:1961} (see also \cite{GilvaryBergstrom:1961}).

It has to be mentioned that the derived laws
Eqs.~(\ref{schumann},\ref{rosin_rammler}) describe the distribution of
fragments only below a certain size (endoclastic vs.~exoclastic
distribution \cite{Gilvary:1961}). For the largest fragments an
equivalently well--defined quantitative statement does not exist. The
deviation between theoretical predictions and experimental
observations typically occurs for grains which together collect about
75--90\% of the total mass \cite{Weichert:1992}.  Moreover, some
diverging theoretical mean values can be understood from the fact that
the assumption of equidistributed flaws on all length scales neglects
the effect of flaw depletion applying to tiny fragments. Flaw
depletion thus sets a lower bound to the range of applicability.
Insofar the size of fragments which is satisfactorily described by the
distributions Eq.~(\ref{schumann}) and Eq.~(\ref{rosin_rammler}) is
restricted to the intermediate range.

When fitting both the more refined Rosin--Rammler law
Eq.~(\ref{rosin_rammler}) and the simplified Schumann law
Eq.~(\ref{schumann}) to diverse experimental data it was found that in
some situations the Rosin-Rammler law proved superior
\cite{Weichert:1992} (mill products \cite{GilvaryBergstrom:1961})
whereas in other cases (specimen fractured in gelatin
\cite{GilvaryBergstrom:1961}) the Schuhmann law was even more
adequate.

Finally, we mention the {\em population balance model}
\cite{Herbstetal:1981} which assumes that the fragment size
distribution can be normalized to the initial particle size,
i.e.~$Q=Q(R_f/R)$. This assumption is based on empirical observations.
Moreover, its central assertion can also be derived when starting from
a Weibull exponent $z=2$ and applying statistical reasoning similar to
the one sketched above.

In our simulations we employed the Rosin-Rammler law
Eq.~(\ref{rosin_rammler}) with $k=R$ and an exponent $\beta=1$. Since
experimental data have shown the mean dimension of the largest
fragment to be roughly 75\% of the original particle size
\cite{GilvaryBergstrom:1961}, i.e.~$R_f^{max}={3\over 4}R$, the
normalized fragment size distribution simply reads
\begin{equation}
  \label{normfragdis}
  Q\left({R_f\over R}\right) =
  {1-\exp\left(-\displaystyle{R_f\over R}\right)\over 
    1-\exp\left(-\displaystyle{3\over 4}\right)}
  ~.
\end{equation}
\subsection{Molecular Dynamics Modelling of Fragmentation}
\label{sec:2.4}
In the present study we exclusively utilize spherical particles
(disks), which means that also the fragments of a disrupted particle
must be spherical. The single fracture statistics, as discussed in
section \ref{sec:2.3}, is put into practice by dismembering a cracking
disk into two fragments. Whereas the size of one fragment is chosen in
accordance with the Rosin-Rammler law the size of the second one is
determined by mass conservation, i.e.~for two-dimensional particles by
area conservation.

\begin{figure}[htbp]
  \centerline{\psfig{figure=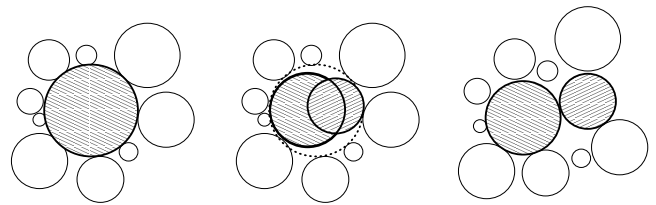,width=8.5cm}}
  \caption{Sketch of the particle fragmentation procedure.
    Immediately after the fragmentation the particles must overlap due
    to mismatch of geometry of the spherical fragments (middle), a
    small repulsive force helps to overcome this unphysical transient
    situation.}
  \label{fig:merkwuerdig}
\end{figure}

In the majority of situations a breaking particle is closely
surrounded by neighbouring ones. Hence, in general, it will be
impossible to place the two fragments avoiding an overlap. However,
since a considerable overlap corresponds to a situation of extreme
compression an unrealistically strong repulsive normal force
(\ref{fnormal}) together with an exceptional deformation energy would
be the consequence. We bypassed this problem by modifying the
interaction between the two fragments in the following way: Up to the
moment of first complete separation of two such fragments $i$ and $j$
the normal force (\ref{fnormal}) at the $k+1$-rst time step is
replaced by
\begin{equation}
  F_{ij}^N=\left\{
    \begin{tabular}{l} 
      $Y\xi_{ij}^* (k)-m_{ij}^{{\rm eff}}\, \gamma_N\, 
      (\dot{\vec{r}}_i - \dot{\vec{r}}_j) \cdot\vec{n}$ 
\\~~~~~~~~~~~~~~~~~~~~~~~if~~~ 
      $(\dot{\vec{r}}_i - \dot{\vec{r}}_j) \cdot\vec{n} < 0$\\
      $\epsilon$ ~~~~~~~~~~~~~~~~~~~~~otherwise\\
    \end{tabular}\right.
\end{equation}
with
\begin{equation}
  \xi_{ij}^* (k)= \left|  \vec{r}_i(k-1)-\vec{r}_j(k-1)\right| - 
  \left|  \vec{r}_i(k)-\vec{r}_j(k)\right|\,.
\end{equation}
This means that the two fragments, on the one hand, resist to further
compression like standard particles and, on the other hand, experience
a small repelling force $\epsilon$ which drives them apart. The
repelling force is chosen sufficiently small in order to suppress
unphysical energy gain of the system; this means that the unavoidable
energy, which due to this force is pumped into the system anyhow, must
be negligible in comparison with the mean kinetic energy of the
grains. After the moment of first complete separation of the two
fragments they also interact in the standard way, as dictated by
Eq.~(\ref{force}). Of course, the interaction of two such fragments
with all other particles is never modified. By trimming of $\epsilon$
we could not only suppress the elastic energy gain but also evidenced
that the period requiring a modified interaction force was
comparatively short. The procedure of fragmentation is sketched in
Fig.~\ref{fig:merkwuerdig}.
We want to mention that multiple fragmentation, i.e. further
fragmentation of the fragments is included in the model and is
frequently observed in simulations.

Let us remark that our algorithm was flexible enough to include
multiple fracture events: By this we mean the observation that a
fragment was apt to break itself even before its first complete
separation from its original twin fragment.

An animated sequence of fragmentation can be found in the
Internet under http://summa.physik.hu-berlin.de/ $\sim$kies/mill/bm.html

Finally, we mention that the proposed algorithm is not the only way to
model particle fragmentation. An alternative algorithm has 
been put forward by {\AA}str\"om and Herrmann
\cite{AstroemHerrmann:1998}.
\subsection{Specification of the Model System}
\label{sec:2.5}
\begin{figure}[htbp]
    \vspace{-0.4cm}
\centerline{\psfig{figure=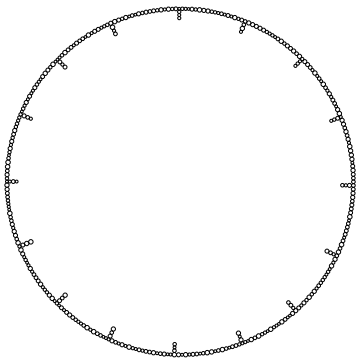,width=5cm}}
    \caption{The cylindrical container is modeled by particles placed
      along the circumference of the container. Roughness is taken
      into account by stochastic variation of the sizes of these
      particles.}
    \label{fig:wand}
    \vspace{-0.2cm}
\end{figure}

As depicted in Fig.~\ref{fig:wand} the cylindrical wall of our system
has been modelled by an ensemble of spheres possessing the same
material constants as the grains inside the container.  However, the
motion of wall spheres is not affected by impacts but, instead,
strictly governed by the continuous rotation of the cylinder.
Moreover, in spite of identical elastic and dissipative constants the
wall particles were considered to be much harder so they never would
break.

A suitable size distribution of wall particles generates the effect of
surface roughness. This type of modelling has been used in several
numerical simulations and is also used in experiments to guarantee
well-defined boundary properties. The wall particles have been placed
along the circumference of the cylinder. Just like in realistic ball
mills we evenly distributed 16 toolbars along the inner periphery of
the cylinder which serve to lift particles and thus prevent them from
sliding downhill. We composed the toolbars again from a collection of
rigid spheres.  As reported in literature
\cite{BernotatSchoenert:1988} the maximum milling efficiency is
achieved for a filling degree, i.e.~the ratio of material and
container volume, of ca.~40\%.

The simulation results to be presented in the next section have been
done with a system consiting in average of about 1000 particles. Due
to fragmentation the number of particles was not constant, i.e., the
particle number fluctuates. The container was modeled by about 500
wall particles as described above.

Initially the spheres which model the grist were randomly positioned
inside the container avoiding any overlap. This artificial initial
condition has to be relaxed by evolution of the system. In this way we
achieved a realistic configuration for any chosen rotation velocity.
Obviously the typical asymptotic configuration depends on the rotation
frequency $\nu$, therefore, we started our numerical monitoring only
after transients had died out. In Fig.~\ref{fig:protos} we present
snapshots of typical configurations 
depending on four different
rotation frequencies.

In technologically relevant ball mills there is a continuous transport
of raw material into the mill and of milled material out of the
device. In most industrial applications this exchange of raw material
and final product is accomplished by axial transport (z-direction) of
material. Of course, with our two-dimensional model we cannot account
for axial transport, therefore, we simulated the material exchange in
continuous operation mode in the following way: Whenever fragments
were generated whose size dropped below a minimum we removed them from
the container. After exact bookkeeping of the mass of removed ``dust''
particles we reinserted a big particle when the accumulated dust mass
exceeded an upper threshold, of course, obeying mass conservation. In
this way we conserved the average filling level throughout the
simulation and, thus, could reach a steady state of the continuous
operation mode.
\section{Simulation Results}
\label{sec:3}
\subsection{Mass distributions}
\label{sec:3.1}
As a first result we studied the frequency dependent fragment size
distribution. This distribution is not necessarily equal to the
distribution of section~\ref{sec:2.3}, which was the result of single
fracture experiments.

It is known~\cite{Kolmogorov:1941,Halmos:1944,Epstein:1947} that for
an arbitrary fracture mechanism, which hierarchically is applied on
all length scales, one will always generate a log-normal
distribution of fragment sizes; this assertion is a pure consequence
of the central limit theorem. However, in our modelling the ball mill
is operated as an open system, i.e.~there is a particle flow into and
out of the mill. Therefore, analyzing the asymptotic size distribution
is also of theoretical interest.

Furthermore, the explicit knowledge of the distribution plays also a
practical role, because from that knowledge one can decide whether a
multi-level fragmentation process is to be preferred to a single-level
one or vice versa. A multi-level process can be either a distribution
of the comminution process to several ball mills, where each is filled
with granular material of a specified narrow size distribution, or a
spatial separation of material of different size in one mill.

Fig.~\ref{fig:fragxxx} shows the differential fragment size
distribution, i.e.~the mass fraction of all particles with size within
a certain interval, for four different rotational frequencies. The
black circles in each column correspond to measurements at different
times and thus reflect the temporal variance. The time period of each
simulation was 10 seconds real time (beyond transients).

\begin{onecolumn}
\begin{figure}[htbp]
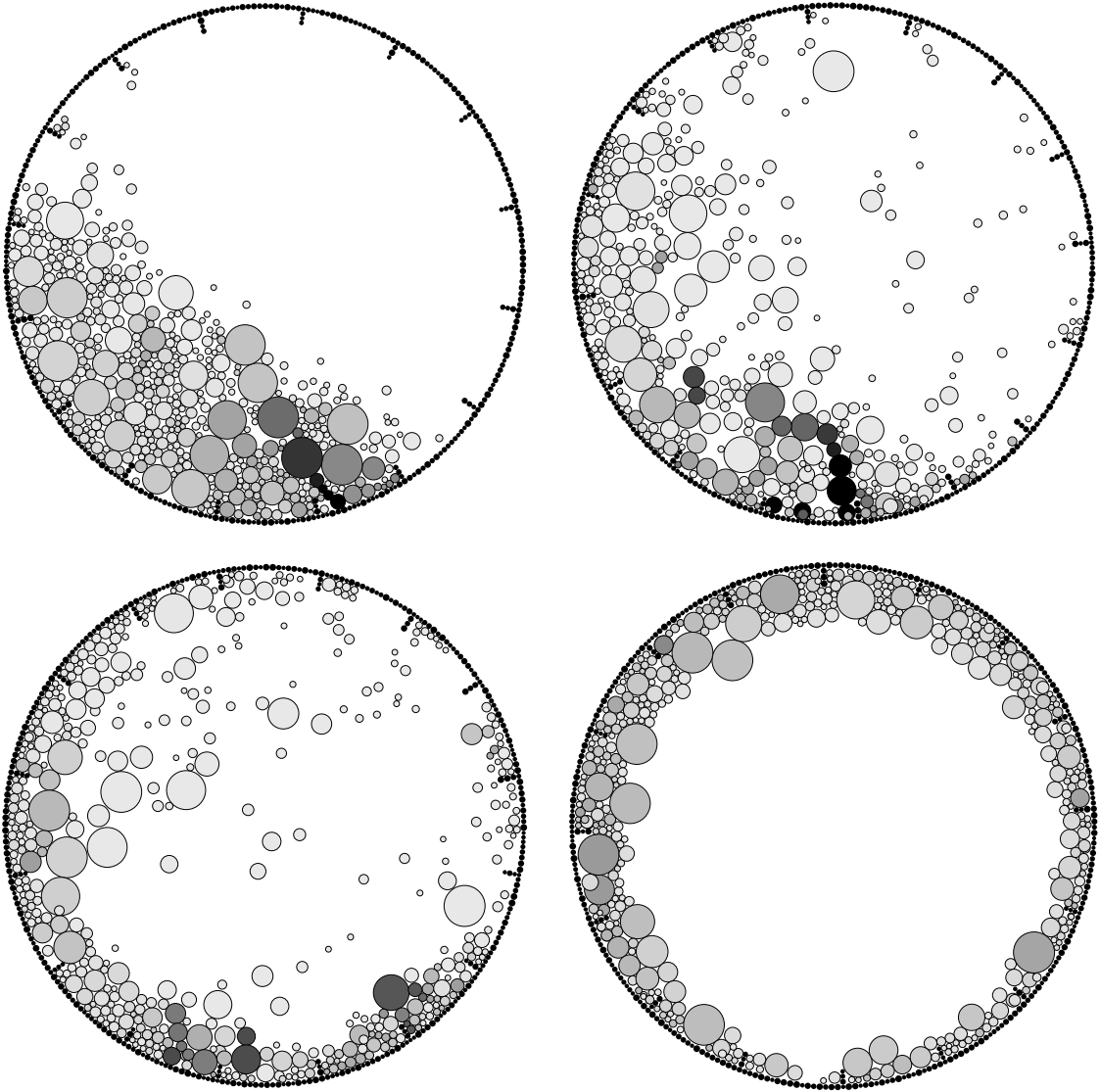

\vspace*{0.2cm}
      \centerline{
        \psfig{figure=figs/proto_0.5_epsBW1,width=7.5cm}
        \hspace{.0cm}
        \psfig{figure=figs/proto_1.0_epsBW1,width=7.5cm}}
      \vspace{0.1cm}
      \centerline{
        \psfig{figure=figs/proto_1.25_epsBW1,width=7.5cm}
        \hspace{.0cm}
        \psfig{figure=figs/proto_2.5_epsBW1,width=7.5cm}}
      \vspace{1cm}
      \caption{Typical dynamical configurations depend on the rotation
        speed: $0.5$\,Hz (top left), $1.0$\,Hz (top right), $1.25$\,Hz
        (bottom left), $2.5$\,Hz. Grey scale codes the instantaneous
        maximum compression (normal force) experienced by each
        particle through contact with its neighbours (light: small
        normal force, dark: large force).  Obviously the largest forces
        occur in the range of intermediate rotation velocities
        ($1.0$\,Hz, $1.25$\,Hz).  Moreover, we identify the location
        of largest forces which can initialize fragmentation to be
        close to the bottom of the rotating cylinder. This important
        observation will be discussed in detail in section
        \ref{sec:4}.}
      \label{fig:protos}
\end{figure}
\end{onecolumn}
\begin{twocolumn}

\begin{figure}[htbp]
  \centerline{\psfig{figure=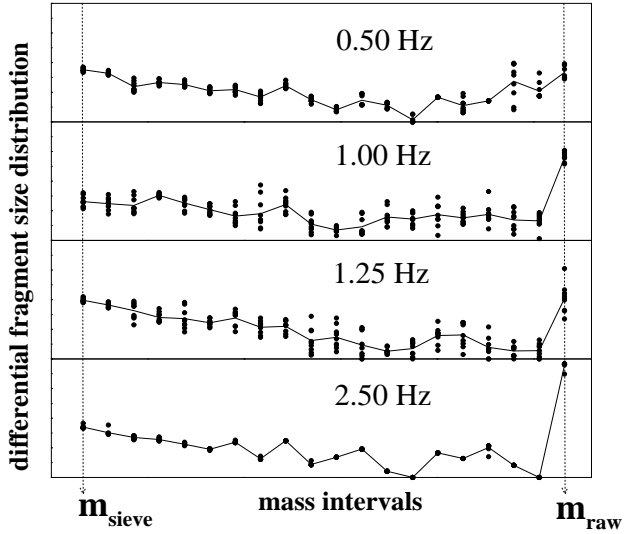,width=9cm}}
  \hspace*{1.5cm}
  \caption{The differential fragment size distribution for different
    rotation velocities. The black circles represent the values for a
    fixed mass interval at different times. The variance of the
    circles illustrates fluctuations of the distribution. The solid
    lines are tracing the averages of each interval. The left border
    of the scale is fixed by the size of the sieve ($m_{\rm sieve}$),
    the right border by the size of the refilled big grains ($m_{\rm
      raw}$).}
  \label{fig:fragxxx}
\end{figure}

Obviously the lowest curve ($2.5$\,Hz) changes only very weakly during
the simulation, which can be interpreted as a low rate of
fragmentation. This curve -- and with restrictions also the top curve
($0.5$\,Hz) -- is determined by the initialization. Unfortunately, the
middle curves are not significantly different, which mainly is a
consequence of the short simulation time (10 seconds). To answer the
interesting questions (theoretical distribution, multi-level process)
it will be necessary to simulate considerably longer time periods.

\subsection{Optimization of the efficiency}
\label{sec:3.2}
In the context of industrial applications the notion of efficiency
mainly focuses on the aspects of maximum comminution rate and,
perhaps, of power consumption. There are many free parameters which
can be varied to maximize the efficiency: speed of rotation, filling
degree, size distribution, cylinder diameter, grist size distribution,
various mill types, etc.

In our analysis we investigated the dependence of the comminution rate
on the speed of rotation. In order to compare our numerical results
with experimental data we normalized quantities in the following way:
The speed of rotation is normalized to the velocity
$n_c=\sqrt{g/M}(2\pi)^{-1}$ for which the centrifugal force balances
the gravitational force whereas the comminution rate is normalized to
its measured maximum. The result of our simulation is plotted in
Fig.~\ref{fig:opti_rate} using filled circles and dashed line.
Clearly, a maximum occurs around $n \approx n_c$ whereas the rate
rapidly diminishes for larger or smaller velocities.
\begin{figure}[htbp]
  \begin{center}
    \centerline{\psfig{figure=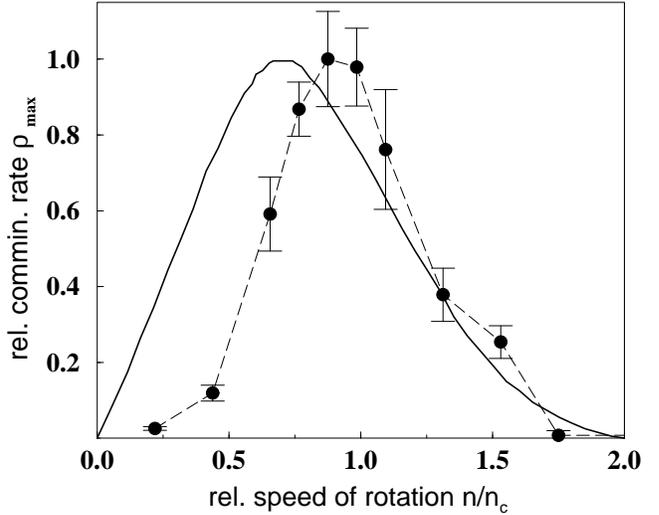,width=8.5cm}}
    \caption{The dependence of the fraction rate on the rescaled
      rotation velocity, simulation (filled circles plus dashed line)
      and experiment (full line) taken from
      ~\cite{BernotatSchoenert:1988} (original~in~\cite{Leluschko:1985}). 
The error bars for the simulation visualize the
      fluctuations during the simulation period of approximately 10
      seconds real time.}
    \label{fig:opti_rate}
  \end{center}
  \vspace*{-0.2cm}
\end{figure}

The full line in Fig.~\ref{fig:opti_rate} depicts a similar curve
following from experiments. The curve shows the relative power
consumption as a function of the normalized speed of rotation.
Qualitatively both curves are in good agreement; systematic deviations
can be explained by the following observation: Strictly speaking both
curves relate to different physical quantities. The fraction rate only
accounts for the energy dissipated through fractures while the
experimental curve includes all dissipative processes, for instance
friction between grains, frictions between the grains and the wall,
dissipative impacts, heating, etc.). As a consequence, the curve for
the simulation should be found below the experimental curve. Moreover,
from the same reasoning it is also clear that observed deviations are
minor if the main portion of the overall dissipated energy results
from the fractures, i.e.~around the maximum.
\subsection{Preferred fragmentation locations}
\label{sec:3.3}
The series of pictures in Fig.~\ref{fig:brech} illustrates the spatial
distribution of fragmentation locations for four different rotation
velocities. We restricted our presentation to the most important lower
right segment of the ball mill (clockwise rotation). To compute this
distribution the coordinates of breaking particles were stored during
the simulation. After the simulation the fraction of all fragmentation
events occurring at places inside little cells (coarse graining) was
computed. In Fig.~\ref{fig:brech} grey values code the frequency: dark
means high frequency and light low frequency respectively.

In the bottom right figure one finds a few fragmentation events near
the cylinder wall. The restriction to medium-grey values results
from the fact, that the rare events are equidistributed across the
inner perimeter.

\begin{onecolumn}
\begin{figure}[htbp]
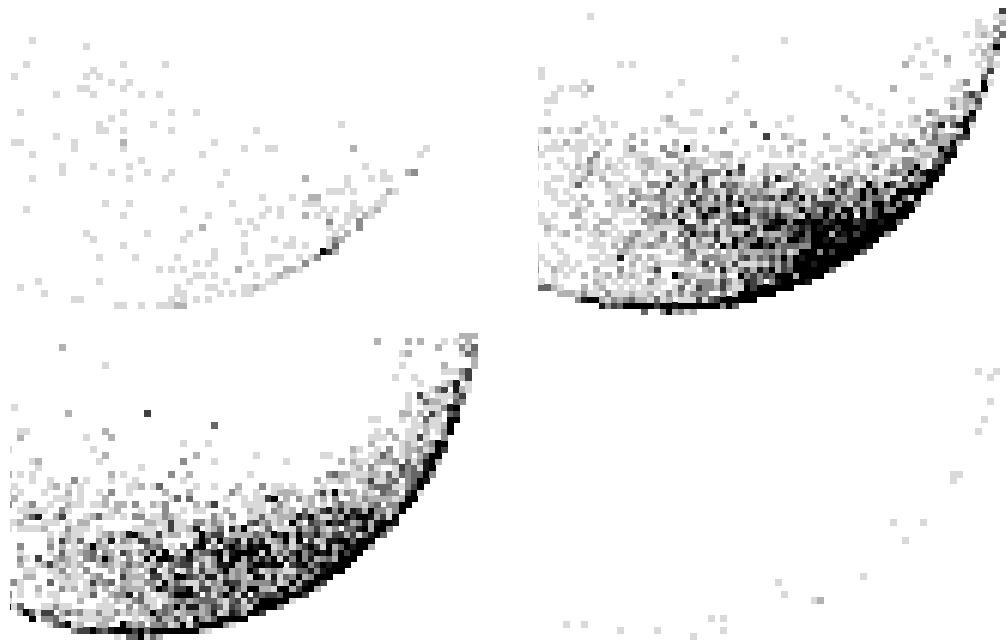

  \vspace*{0cm}
  \centerline{
    \psfig{figure=figs/loc_brech_0.5_epsBW,width=6.5cm,bbllx=130pt,bblly=1pt,bburx=400pt,bbury=180pt,clip=}
    \hspace{0.3cm}
    \psfig{figure=figs/loc_brech_1.0_epsBW,width=6.5cm,bbllx=130pt,bblly=1pt,bburx=400pt,bbury=180pt,clip=}}
  \vspace{0.0cm}
  \centerline{
    \psfig{figure=figs/loc_brech_1.25_epsBW,width=6.5cm,bbllx=130pt,bblly=1pt,bburx=400pt,bbury=180pt,clip=}
    \hspace{0.3cm}
    \psfig{figure=figs/loc_brech_2.5_epsBW,width=6.5cm,bbllx=130pt,bblly=1pt,bburx=400pt,bbury=180pt,clip=}}
  \caption{Spatial distribution of fragmentation locations for
    different rotation velocities: $0.5$\,Hz, $1.0$\,Hz, $1.25$\,Hz,
    $2.5$\,Hz (from top left to bottom right). Grey values code the
    frequency of fragmentation in a certain region (cell), dark
    meaning high frequency and light low frequency respectively.
    Preferred fragmentation regions are found near the bottom of the
    mill, i.e.~deep inside the granular material. This observation is
    in good agreement with experimental results and the results from
    chapter~\ref{sec:4}.}
    \label{fig:brech}
\end{figure}
\begin{figure}[htbp]
\centerline{\psfig{figure=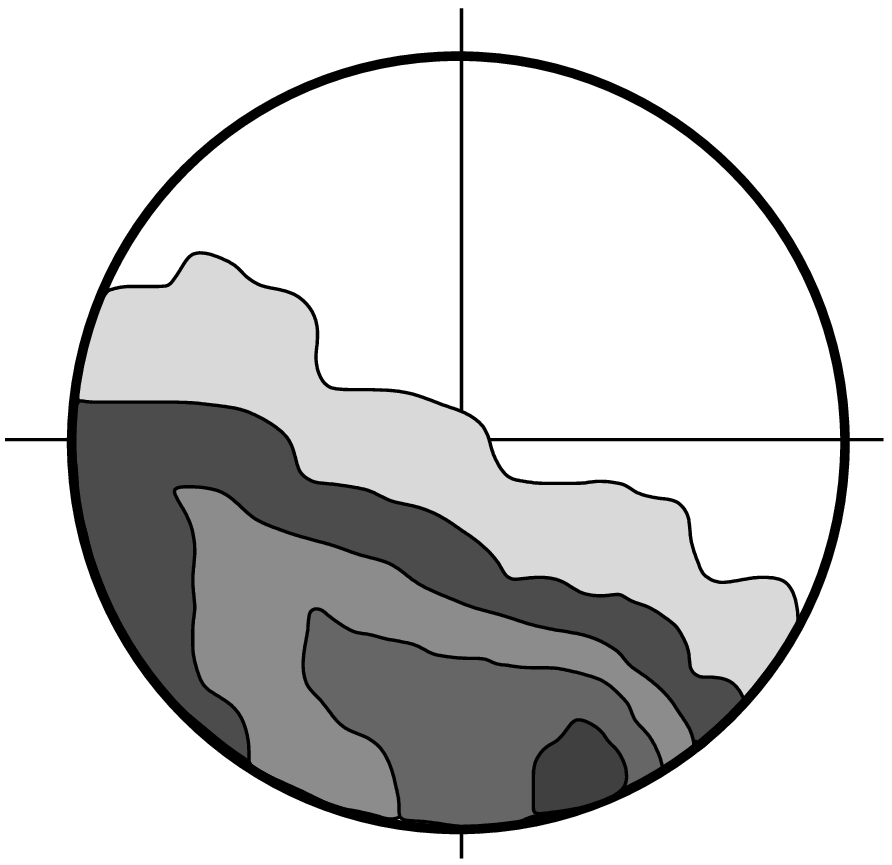,width=5.5cm}
\psfig{figure=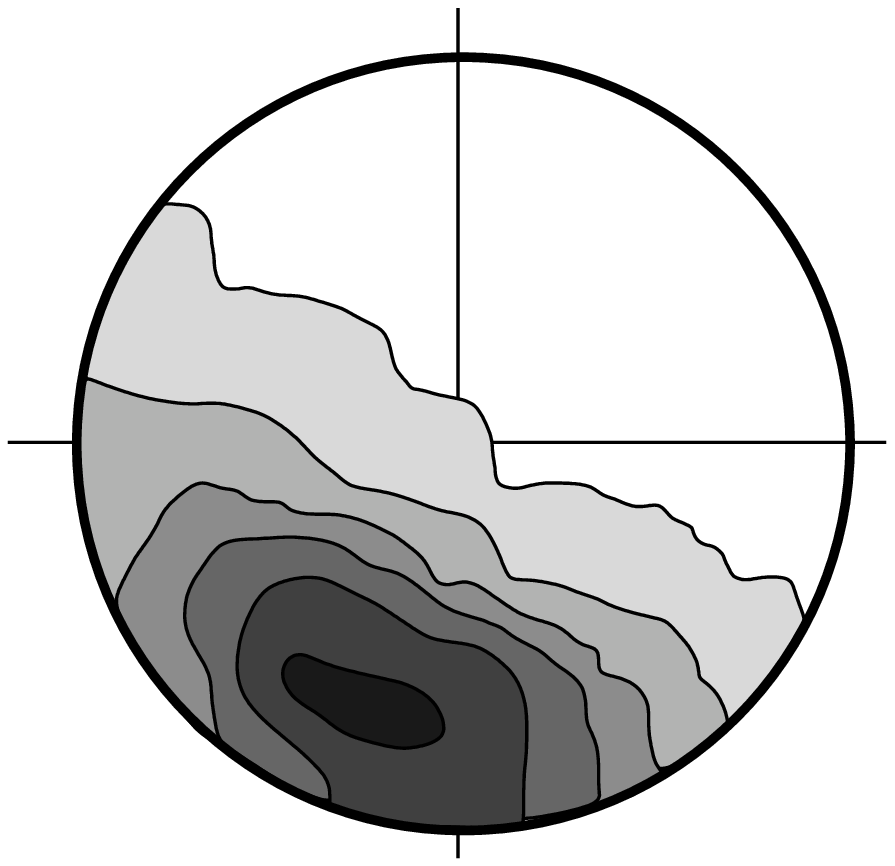,width=5.5cm}
\psfig{figure=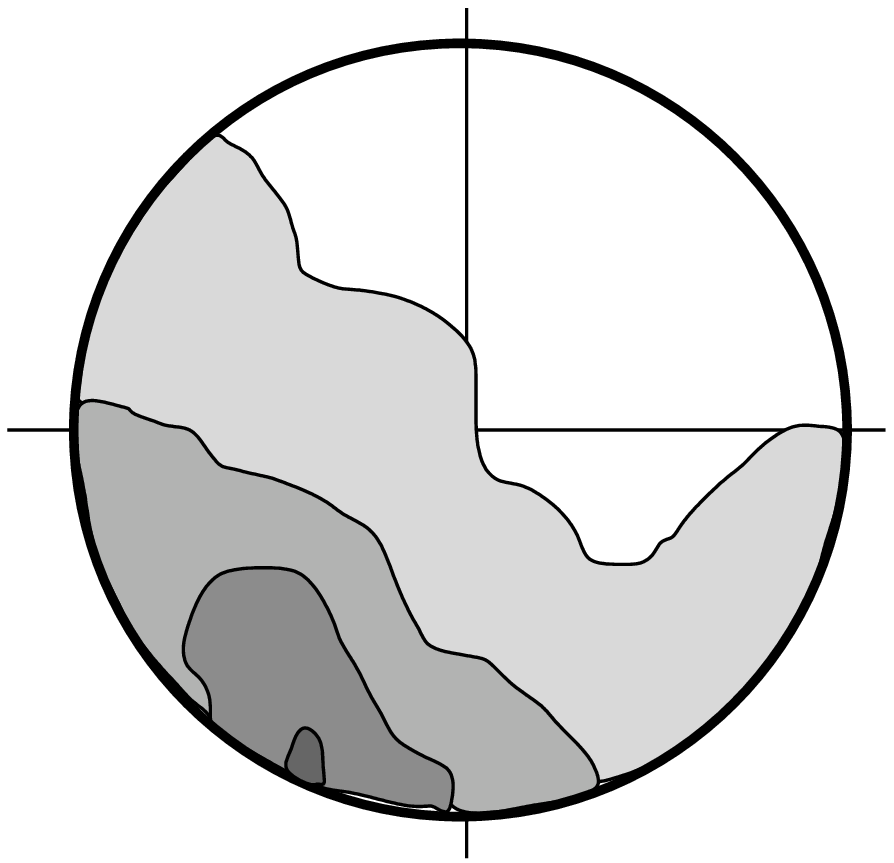,width=5.5cm}}
  \caption{Spatial distribution of the strain for different rotation 
    velocities and filling grades. Again, grey levels code strain
    values. The regions of maximal strain are located inside the
    material. This figure is redrawn from ref.~\cite{Rothkegel:1992}.}
  \label{volk:ing}
\end{figure}

\end{onecolumn}
\begin{twocolumn}
\end{twocolumn}

The predominance of rather light grey (or even white) pixels in the
rest of the figures hints at regions where fragmentation events are
rare as compared to the sparse medium-grey to dark pixels which
indicate regions of high fracture probability. The observation that
most fragmentation events occur deep inside the material and not near
the surface, which is heavily agitated by impacts, is, at first sight,
somewhat counter-intuitive.  An explanation for this fact will follow
from a closer analysis of the local pressure concentration in the next
section.

\section{Spatial distribution of pressure in ball mills}
\label{sec:4}
In this chapter we will dwell on the physical origin of the unusual
pressure distribution~\cite{pressure} measured in experiments.
  
  
  
  

As can be concluded from Figs.~\ref{fig:brech} and \ref{volk:ing}, the
majority of fragmentation events does not occur near the free surface
of the granular material but deep inside the material near to the
walls. This strange behavior was observed already in experiments by
Rothkegel and Rolf~\cite{Rothkegel:1992,Rolf} in two-dimensional model
mills. They used instrumented balls which emitted a flash of light
whenever the strain exceeded a fixed threshold. The positions of
recorded flashes were digitized and from the statistics of these
data they inferred the stationary strain distribution inside the ball
mill. For various threshold values they always found the peak strain
deep inside the material (see~Fig.~\ref{volk:ing}). This was
surprising because inside the material relative grain velocities are
rather low in comparison with relative velocities at the surface.

This experimental finding, together with its validation by our
simulation (see the previous section), calls for an explanation. We
will solve this challenging, so far open question using the
powerful possibilities of molecular dynamics, namely the option to
monitor physical quantities which are hardly accessible through
experiments.

In the following simulations we ignored fragmentation events -- which
are not relevant for our explanation -- simply for the sake of
computational efficiency, so we could afford longer simulation time.
This time the radius of the two-dimensional ball mill was set to
$M=4$\,cm. The mill was filled with $N=800$ spherical grains with
radii equally distributed in the interval $[0,05,0,11]$\,cm. We
performed simulations with three different rotation velocities
$\Omega_{\rm I} = 2$\,Hz, $\Omega_{\rm II} = 10$\,Hz and $\Omega_{\rm
  III} = 19$\,Hz.

Fig.~\ref{volk:snaps} shows snapshots of our performed simulations.
The pictures in the top row are taken form a simulation with rotation
velocity $\Omega = \Omega_{\rm I}$. The velocity is sufficiently
enough to observe a continuous flow at the surface. The surface shape
is similar to a plane. The regime with discontinuous flow (for lower
rotation velocities) will not be investigated in this paper, because
it is not relevant for technical purposes. For a detailed discussion
of the interesting observations at the transition between continuous
and stick-slip flow we refer to the work by
Rajchenbach~\cite{Rajchenbach:1990}.

The picture in the middle row of Fig.~\ref{volk:snaps} depicts
snapshots of simulation with $\Omega = \Omega_{\rm II}$. The surface
of the material no longer shapes a plain but the grains fall down a
steep slope hitting a flat surface.

The bottom row of Fig.~\ref{volk:snaps} exhibits the snapshot of a
simulation with the highest rotation velocity $\Omega=\Omega_{\rm
  III}$. Here the grains are carried away with the fast rotation of
the cylinder. Beyond a certain point they are likely to loose contact
with the wall and follow a free parabola before impacting back down on
the surface.

Moreover, the snapshots of Fig.~\ref{volk:snaps} provide an intuitive
understanding, what mechanism is responsible for the pressure
distribution mentioned above. Again, grey values in these pictures
codes the local pressure $P_i$ acting on grain $i$ given by
\begin{equation}
  P_i = \sum\limits_j F_{ij}^N~.
  \label{pressure}
\end{equation} 
The index $j$ is running over all particles which are in contact with
grain $i$. Dark circles denote high pressure, light spheres low
pressure.

Obviously many of the heavier stressed particles are deep inside the
material, which is in good agreement with the experimental observation
by Rothkegel and Rolf~\cite{Rothkegel:1992,Rolf}. Most importantly, the
pictures reveal that grains experiencing peak stress often form linear
structures. In the following we focus on the development and the
properties of such force chains. By numerical analysis we will prove
that those force chains are the crucial physical reason for the
observed pressure distribution.

We defined force chains by a set of three self-evident conditions:
Grains $i$, $j$ and $k$ are considered to be members of the same force
chain if:
\begin{enumerate}
\item Particles $i$, $j$ and $j$, $k$ are next neighbours.
\item The pressure acting on each of the grains exceeds a certain threshold
  ($P_i > 1000$\,g\,cm\,s$^{-2}$).
\item The connecting lines between $i$, $j$ and $j$, $k$ form an angle
  larger than $150^o$, i.e.~the centres of three grains almost fall
  on a line.
\end{enumerate}
These three conditions are evaluated by a computer algorithm. In
Fig.~\ref{volk:snaps} all grains belonging to a force chain are
graphically connected by lines.

We see that most of the harder stressed grains could be assigned to a
force chain. From this we conclude that the main part of the static
and dynamic pressure propagates along the force chains. The
pressure acting on a highly stressed grain inside a force chain is up
to 100 times larger as compared to the average pressure of the
neighbouring grains not belonging to a force chain, i.e.~the force
distribution is strongly inhomogeneous.

The development of such force chains is not restricted to grains in
a ball mill but has been observed in different granular systems by
various
experimentalists~\cite{DuranMazoziLudingClementRajchenbach:1996,Behringer:1993,LiuNagelSchecterCoppersmithMajudarNarayanWitten:1995,Dantu:1967}
and
numerically~\cite{Schoellmann:1999,EsipovPoeschel:1997,Wolf:1996}.
It could be shown, that one of the granular phases is characterized
by the observation of force chains~\cite{EsipovPoeschel:1997}. From
this one might conclude that the occurrence of force chains is an
inherent feature of all granular matter, which may substantially
contribute to its specific characteristics.

\begin{onecolumn}
\begin{figure}[htbp]
  \vspace*{0.5cm}
  \centerline{\psfig{figure=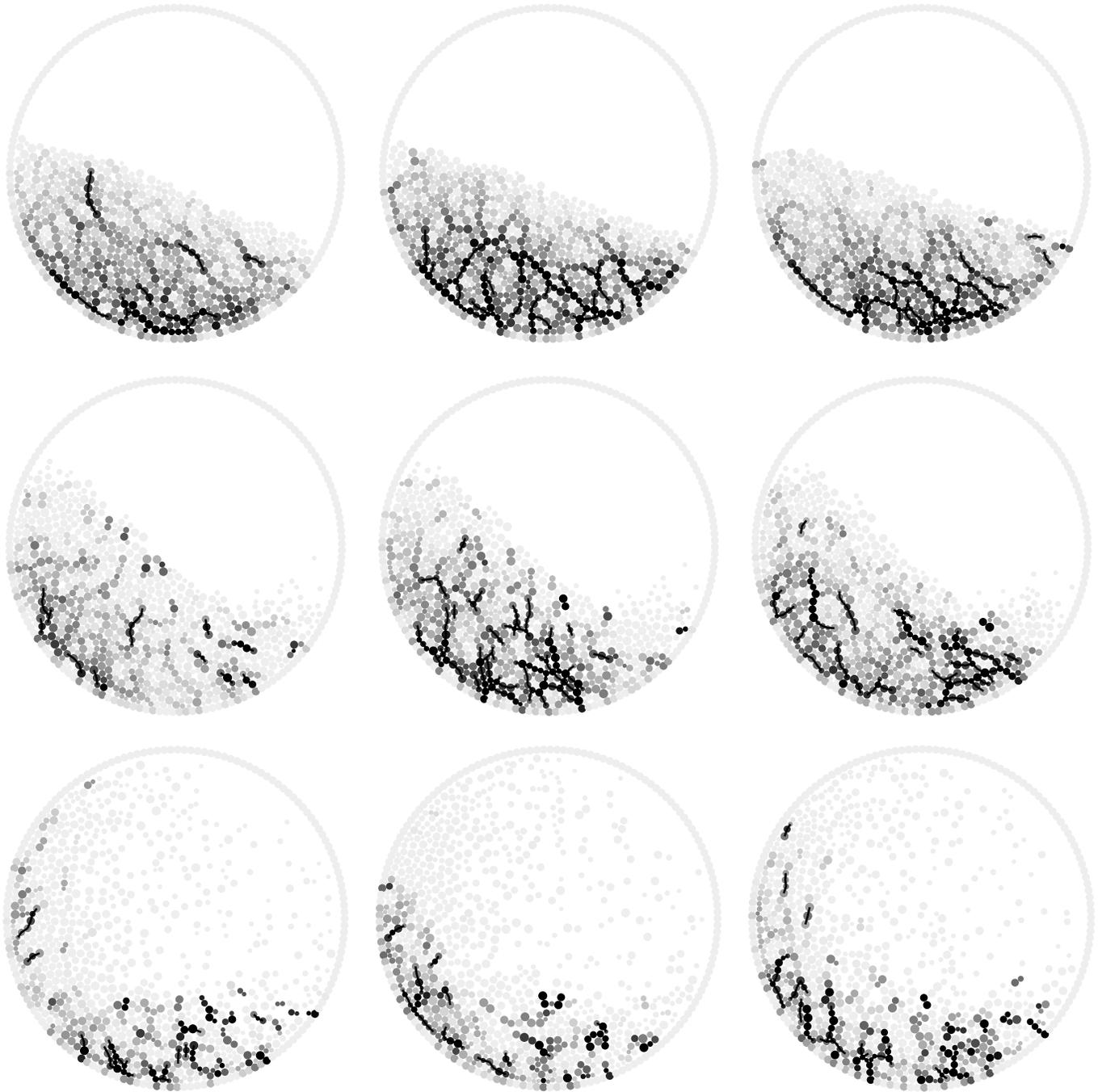,width=18cm}}
  \vspace*{1cm}
  \caption{Snapshots of the simulation for different rotation
    velocities. Grey values code the local pressure $P_i$ (see text).
    Linear segments connect grains belonging to a force chain.
    $\Omega_{\rm I} = 2$\,Hz (top), $\Omega_{\rm II} = 10$\,Hz
    (middle), $\Omega_{\rm III} = 19$\,Hz (bottom).}
\label{volk:snaps}
\end{figure}
\end{onecolumn}
\begin{twocolumn}
\end{twocolumn}

\begin{figure}[htbp]
  \vspace*{0.5cm}
  \centerline{\psfig{figure=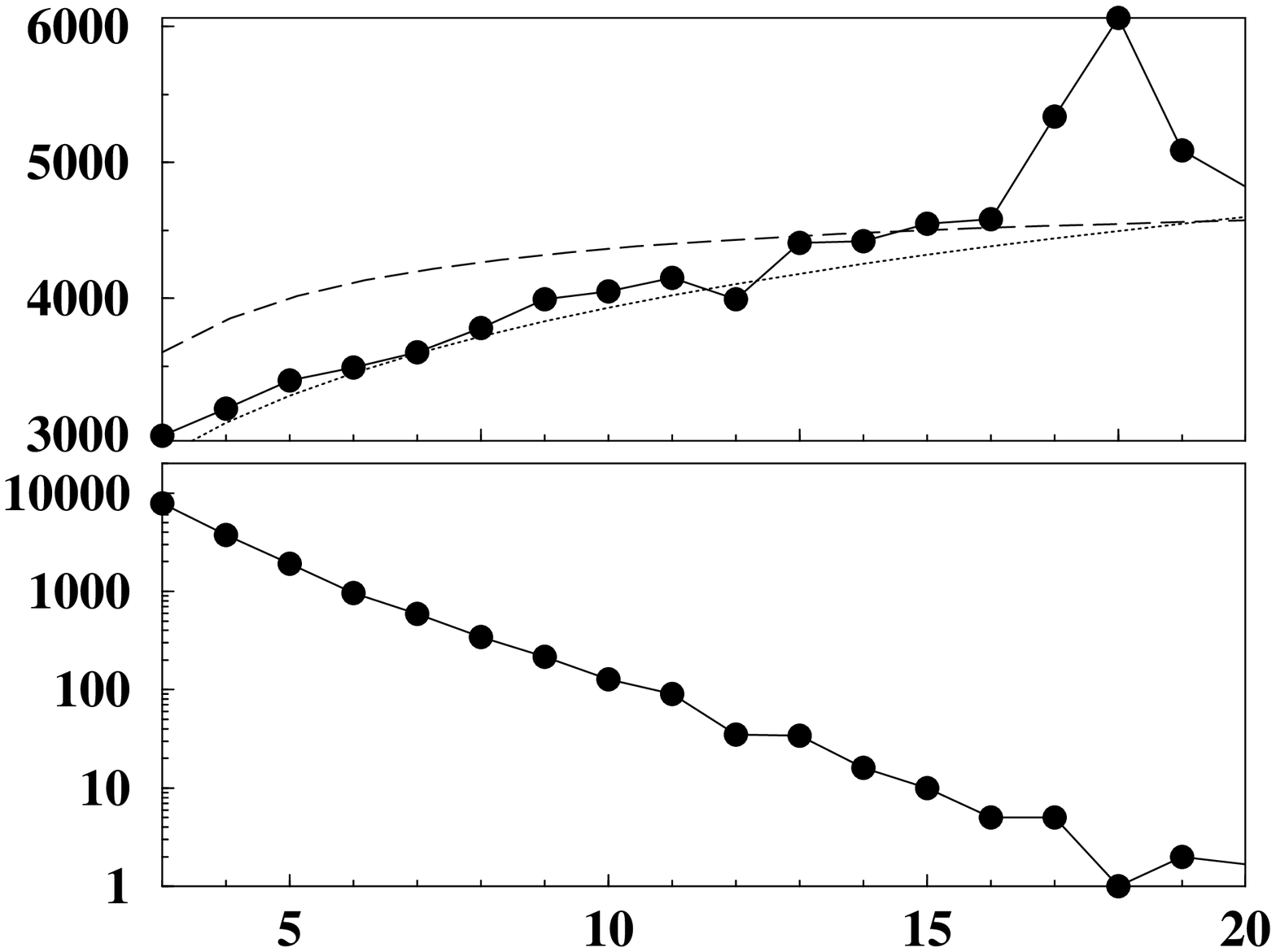,width=7.6cm}}
  \vspace*{0.2cm}
  \centerline{\psfig{figure=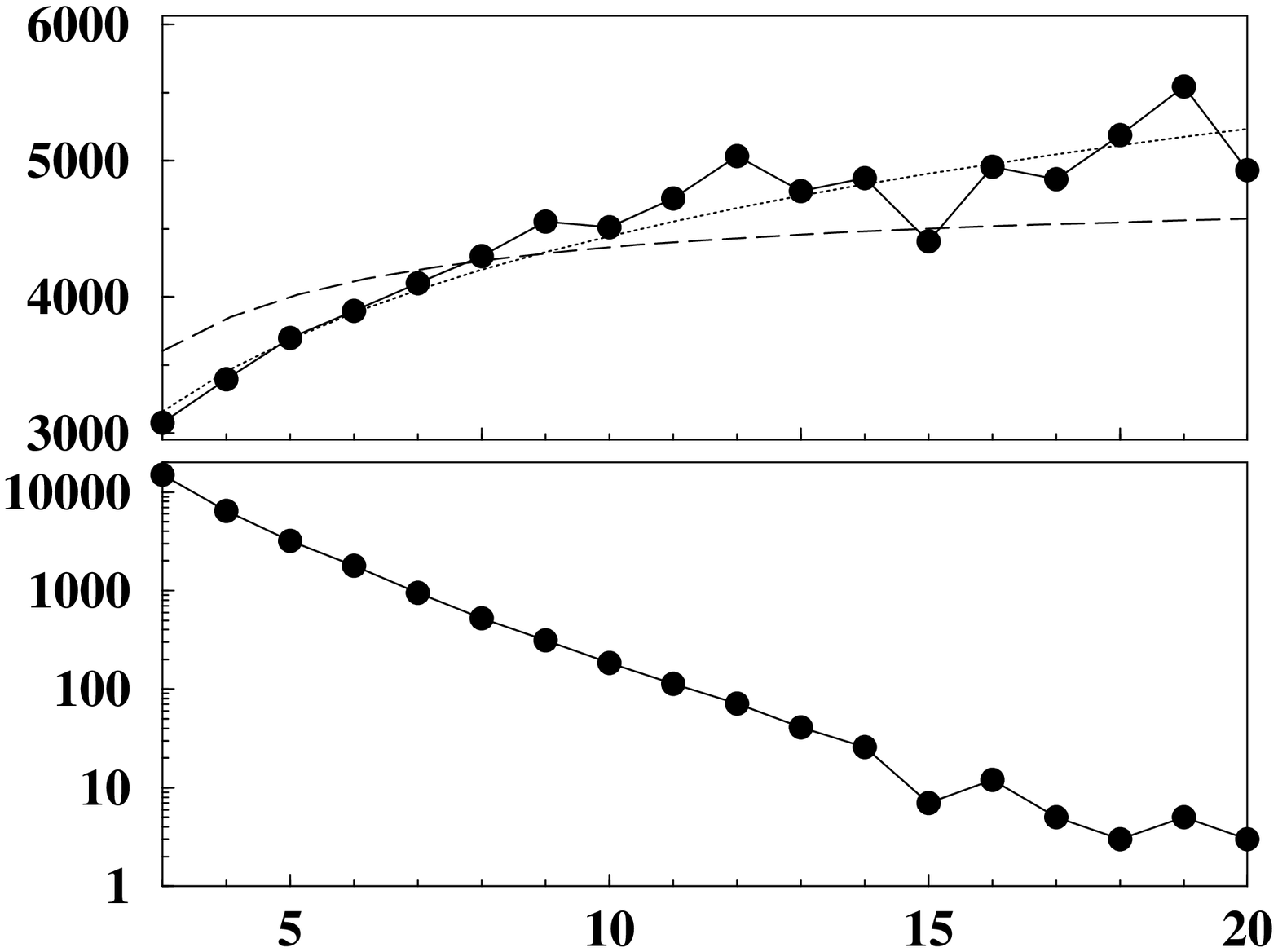,width=7.6cm}}
  \vspace*{0.2cm}
  \centerline{\psfig{figure=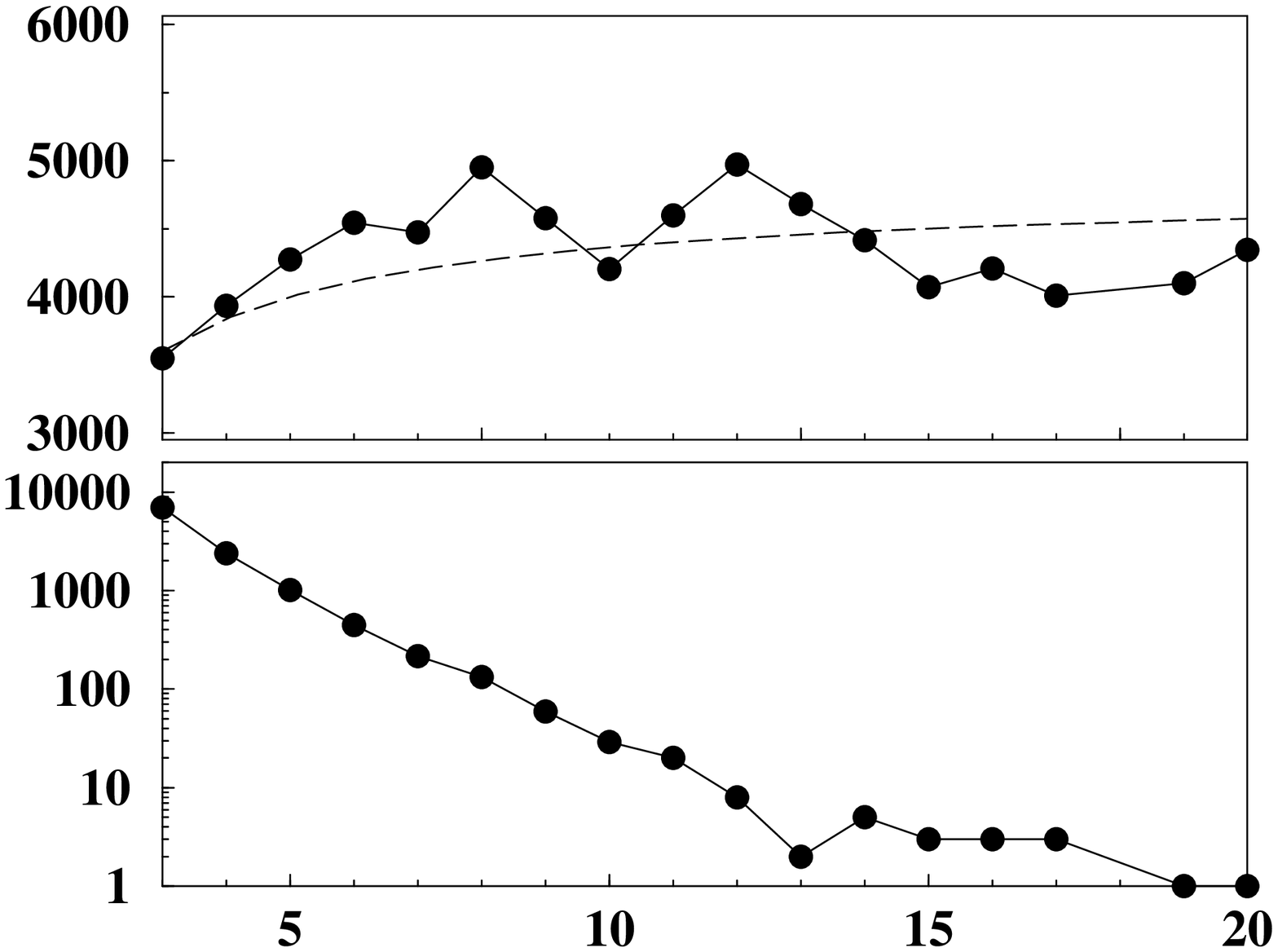,width=7.6cm}}
  \vspace*{0.5cm}
  \caption{Within each pair of the triplet (I,II,III) the lower plot
    presents the frequency distribution whereas the upper one shows
    the maximal pressure inside a force chain both as a function of
    its length $L$ (abscissa for all plots).  $\Omega_{\rm I} = 2$\,Hz
    (top), $\Omega_{\rm II} = 10$\,Hz (center), $\Omega_{\rm III} =
    19$\,Hz (bottom) (explanation of lines see text).}
\label{volk:props}
\end{figure}

Fig.~\ref{volk:props} comprises six plots which come in three pairs;
the lower plot of each pair always depicts the frequency distribution
of force chains whereas the related upper one presents the
(statistically evaluated) maximal pressure both as a function of the
length $L$ of a force chain (which is nothing but the number of
particles belonging to the chain). These pairs are shown for three
different rotation velocities varying from $\Omega_{\rm I}$ (top) to
$\Omega_{\rm III}$ (bottom).

The frequency of a force chain drops exponentially with increasing
length $L$. For the highest rotation velocity ($\Omega_{\rm III}$) the
statistics becomes unreliable for chains longer than $L=13$. This
behaviour, which is in contrast to observations for slower measured
rotations, is due to the higher energy input, which causes a
decompactification of the material.

Concerning the earlier mentioned spatial strain distribution the {\em
  maximal} pressure measured in each force chain is of paramount
interest rather than the {\em average} pressure in a force chain (see
section~\ref{sec:2.2}). For lower rotation velocities $\Omega_{\rm I}$
and $\Omega_{\rm II}$ a monotonous increase of the maximal pressure
with the length $L$ can be observed. For a discussion of these curves
one has to take into account that the pairs of plots in
Fig.~\ref{volk:props} do {\em not} constitute independent quantities.
Choosing a sample of $L$ particles from an ensemble of particles on
which forces with a given distribution are acting, the maximal force
in this sample will increase with increasing $L$. Therefore, in order
to prove significance, the result in Fig.~\ref{volk:props} has to be
weighted against this purely statistical effect.

The probability to measure a maximal pressure $P^{\rm max}$ in the
interval $P^{\rm max} \in [x;x+dx]$ in a sample of size $L$ is equal
to the probability that all particles feel a pressure $P<x+dx$ minus
the probability that all $L$ particles feel a pressure $P<x$.
\begin{equation}
  p_L(P^{\rm max} \in [x;x+dx]) = 
  \left[p(P<x+dx)\right]^L - \left[p(P<x)\right]^L
  \label{eq.WKS}
\end{equation}

Assuming equally distributed pressures $P \in [0:P_m]$, $p(P<x)$ means
\begin{equation}
  p(P<x) = \int\limits_0^x \frac{1}{P_m} dx' = \frac{x}{P_m} ~.
  \label{eq.wkconst}
\end{equation}

By inserting Eq.~(\ref{eq.wkconst}) into Eq.~(\ref{eq.WKS}), one finds
in first order of $dx$
\begin{equation}
  p_L(P^{\rm max} \in [x;x+dx]) = 
  \frac{1}{\left(P_m\right)^L} L \, x^{L-1} dx~.
\end{equation}

Therefore, the average maximal pressure as a function of sample size
$L$ is given by
\begin{eqnarray}
  \left<P^{\rm max}(L)\right> & = & 
  \frac{1}{\left(P_m\right)^L} \int\limits_0^{P_m} x 
  \, L \, x^{L-1} dx\nonumber \\
  & = & \frac{1}{\left(P_m\right)^L}
  \frac{L}{L+1} P_m^{L+1}\nonumber \\
  & = & \frac{L}{L+1} P_m~.
\end{eqnarray}
The dashed line in Fig.~\ref{volk:props} shows this statistical effect
for $P^{\rm max}$ with $P_m = 4800 g$\,cm\,s$^{-2}$. Obviously, these
curves do not reach up to our numerical data of the top and center
figure; especially for larger $L$ the dashed line falls far below. The
numerical data points in the bottom figure (high rotation velocity) to
some extent coincide with the dashed curve, which only represents the
statistical effect discussed above. Hence, we can conclude that for
lower velocities the maximum stress significantly concentrates within
long force chains whereas for high rotation velocities a significant
effect of force chains is not evident.

From experimental and numerical results (for instance
\cite{LiuNagelSchecterCoppersmithMajudarNarayanWitten:1995,Howell:1997,Radjai:1996})
it is known, that the observation of force chains is connected to a
strongly inhomogeneous pressure distribution. The distribution is
essentially an exponentially decreasing function with increasing
pressure. A simplified expression for this distribution is given by
\begin{equation}
  p(P) = a \, \exp\{-a \, P\}
\end{equation}

In analogy to Eq.~(\ref{eq.wkconst}) the probability to measure a
pressure $P<x$ is given by
\begin{equation}
  p(P<x) = \int\limits_0^x a \, 
  \exp\{-a \, x^\prime\} dx^\prime = 1 - \exp\{-a \, x\}~.
  \label{eq.wkexp}
\end{equation}
Inserting this into Eq.~(\ref{eq.WKS}) we obtain in the first order of
$dx$
\begin{eqnarray}
&&  p_L(P^{\rm max} \in [x;x+dx])=\nonumber\\ 
&&~~=  \left(1-e^{-a(x+dx)}\right)^L - 
  \left(1-e^{-a \, x}\right)^L\nonumber\\
  &&~~= L (1-e^{-a \, x})^{L-1} e^{-a \, x} a \, dx~.
\end{eqnarray}

The average of the maximal pressure can be calculated by integrating
\begin{eqnarray}
&&  \left<P^{\rm max}(L)\right> = 
  \int\limits_0^\infty x \, L \left(1-e^{-a \, x}\right)^{L-1} 
  e^{-a \, x} a \, dx\nonumber\\
&&=  \lim_{N\to\infty}\! \left\{\! x  \left.\left(1-e^{-a \, x}\right)^L 
    \right|_0^N - \int\limits_0^N\! \left(1-e^{-a \, x}\right)^L dx\right\}.
\end{eqnarray}
An iterative splitting 
\begin{equation}
(1-e^{-a \, x})^i = (1-e^{-a \,  x})^{i-1} \, (1-e^{-a \, x})
\end{equation}
 results in a summation of integrals of
the type 
\begin{equation}
\int\limits_0^N (1-e^{-a \, x})^{i-1}\, e^{-a \, x} dx =\frac{1}{a\,i}\,. 
\end{equation}
Thus, we obtain
\begin{equation}
  \left<P^{\rm max}(L)\right>  =  \sum\limits_{i=1}^L \frac{1}{a \, i}~.
\end{equation}

A restriction of the statistics to values of the pressure
$P>1000$\,g\,cm\,s$^{-2}$, performing an analogous calculation,
results in
\begin{equation}
  \left<P^{\rm max}(L)\right>  =  
  \sum\limits_{i=1}^L \frac{1}{a \, i} + 
  1000\mbox{\,g\,cm\,s}^{-2}\,.
\end{equation}
From the simulations we extracted the pressure distribution yielding
-- as expected -- an exponential decrease. The prefactor is $a=0.0009$
for $\Omega = 10$\,Hz and $a=0.001$ for $\Omega = 2$\,Hz. The
corresponding curves are plotted in Fig.~\ref{volk:props} as dotted
lines and reasonably describe to the measured values. Thus, we
indirectly can conclude from the measurement that we have an
exponential distribution, which indicates the occurrence of force
chains.

\begin{figure}[htbp]
  \centerline{\psfig{figure=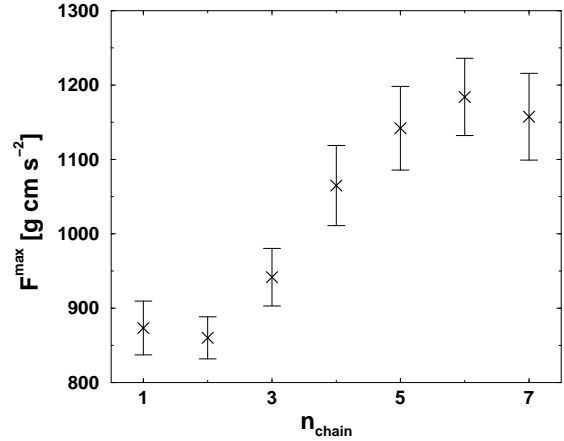,width=7.2cm,angle=270}}
  \caption{The maximal force acting on a particle as a function of 
    the relative position in the force chain. The monotonous behavior
    indicates that forces are accumulating downwards along the
    chain.}
  \label{fig:monoton}
\end{figure}
Further insight in the mechanism of force chains is achieved through
Fig.~\ref{fig:monoton} which shows the average of the maximal force
acting on a grain as a function of its position within the force
chain. The average was performed for all force chains of length $L=8$.
The top particle of a chain, corresponding to the maximal
$y$-coordinate, was defined as position $1$. The monotonous increase
of this maximal force indicates that, on the average, strain
accumulates downwards along the force chain. In this respect a force
chain is similar to a beam in a framework: The masses and momenta of
all particles above a force chain are supported by this chain, with
the consequence that lower particles of the chain are heavier loaded
than upper ones. This trend is clearly visible in the data of
Fig.~\ref{fig:monoton}. Furthermore, a force chain can shield
neighbouring particles, i.e.~these grains experience a much weaker
force which can be observed in the simulation data as well.

Figure~\ref{volk:ldist} shows the spatial pressure distribution for
different rotation velocities. Here grey values code the {\em
absolute} local pressure, with light pixels indicating low and
dark pixels high values, in order to allow for a direct comparison
of all three images.

The presented data are temporal averages. To extract the spatial
distribution of this data, the two-dimensional space was
coarse-grained. For each time step the positions of all particles were
mapped to the grid and the resulting instantaneous local pressure was
added to its related cell. Notice that the averaged data does not
reflect the occurrence of force chains.

\begin{onecolumn}
\begin{figure}[htbp]
  \vspace*{0.5cm}
  \centerline{
    \psfig{figure=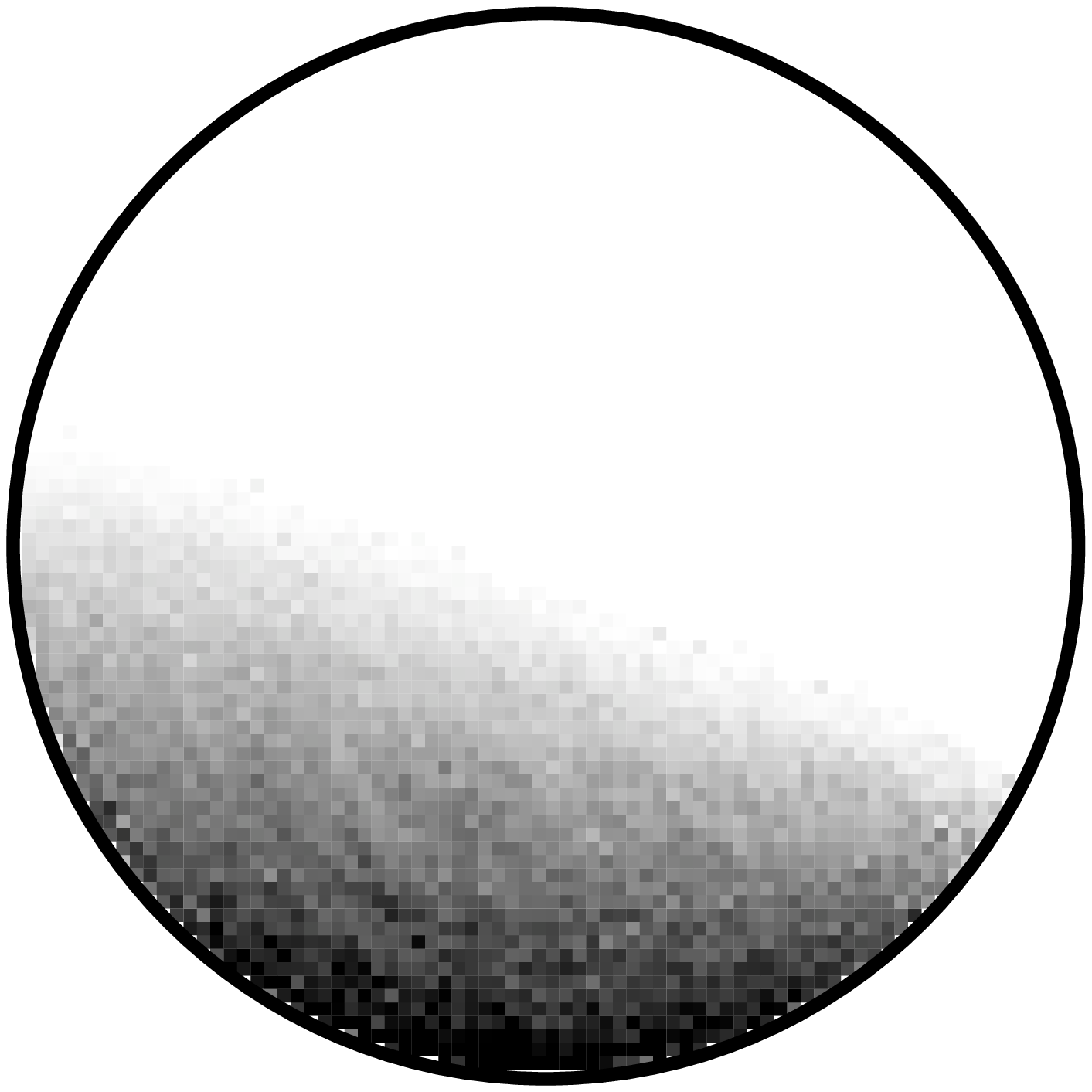,width=5.5cm}
    \psfig{figure=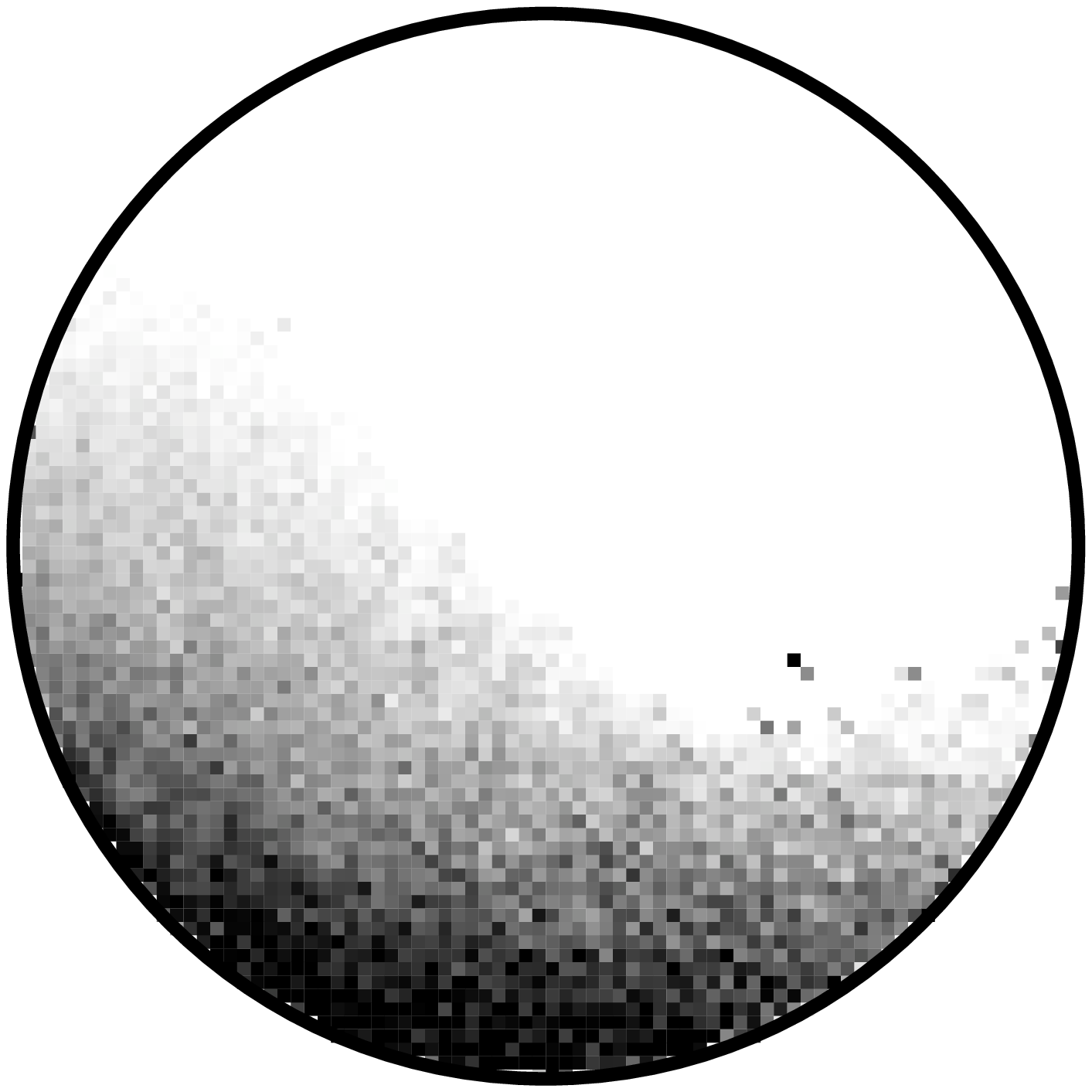,width=5.5cm}
    \psfig{figure=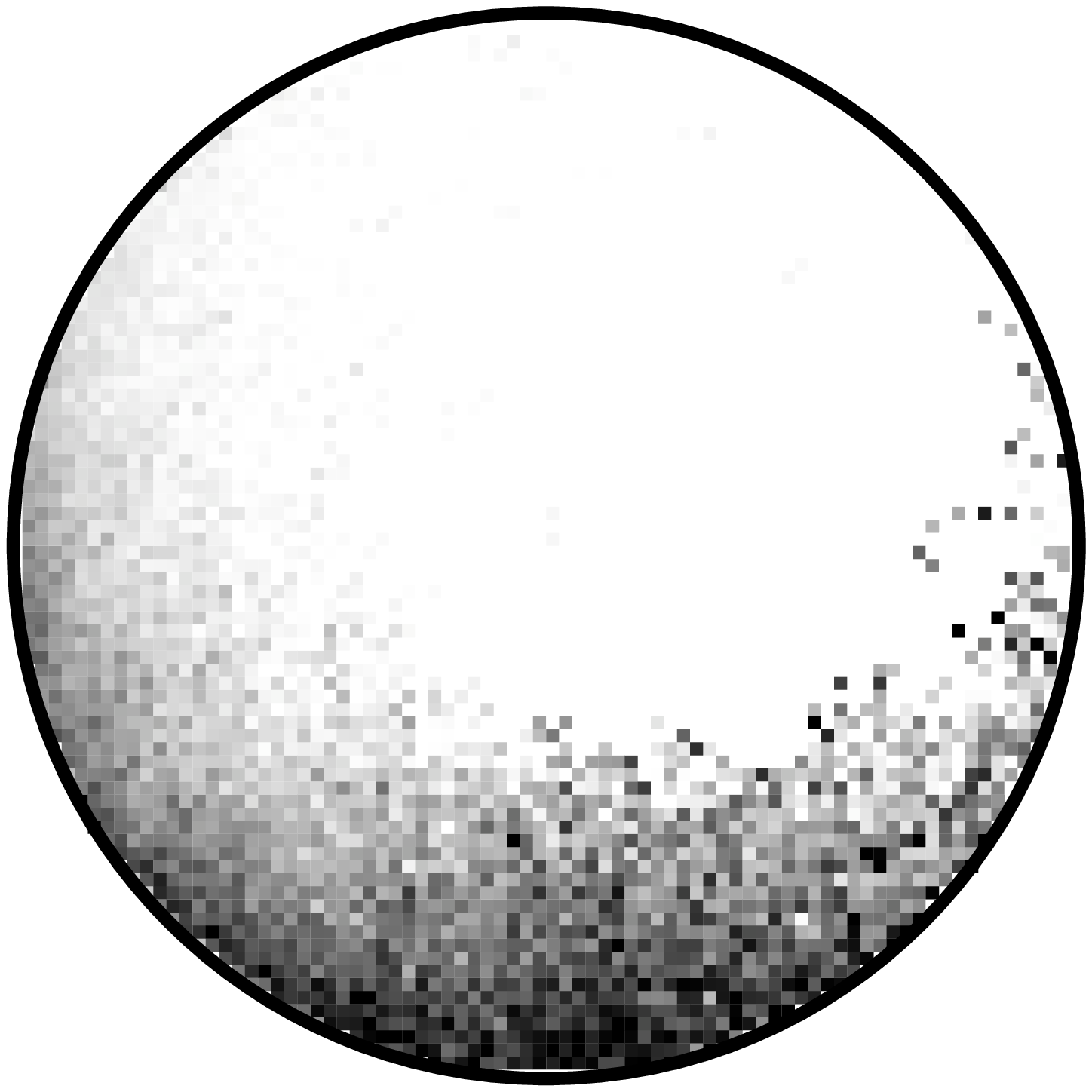,width=5.5cm}}
  \vspace*{1cm}
  \caption{Spatial distribution of the pressure. Grey values code the 
    absolute pressure, with light pixels indicating low and dark
    pixels high pressure values. The grey scale is unique for all
    three figures to allow for a direct comparison. From left to
    right: $\Omega = 2$\,Hz, $\Omega = 10$\,Hz, $\Omega = 19$\,Hz.
    While for lower rotation velocities the maximum is deep inside the
    material near to the wall, for higher rotation velocity it is near
    to the free surface, which is heavily agitated by impacts.}
  \label{volk:ldist}
\end{figure}
\begin{figure}[htbp]
  \vspace*{0.5cm}
  \centerline{\psfig{figure=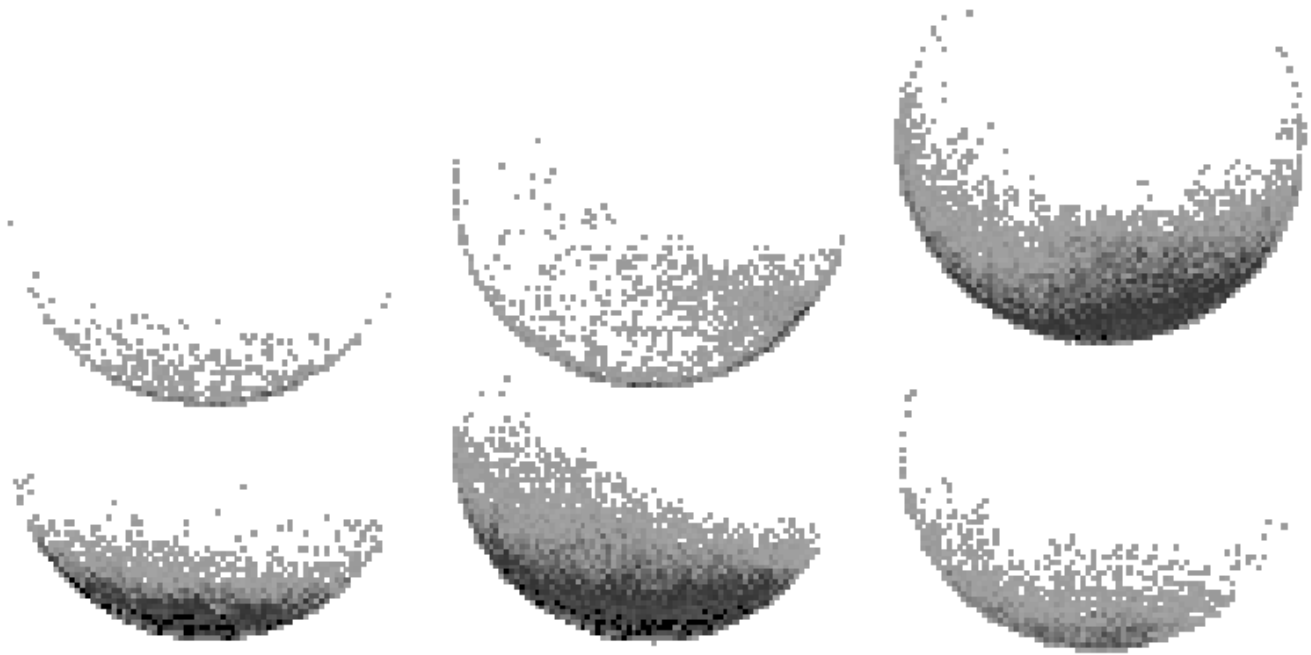,width=17.5cm}}
  \vspace*{1cm}
  \caption{Spatial frequency distribution of particles, exerted to a 
    pressure $P > 3000$ g\,cm\,s$^{-2}$. Grey values, unique
    throughout the images, code the frequency: dark pixels high
    frequency and light pixels low frequencies. The top row only
    accounts for particles which do not belong to a force chain while,
    in contrast, the bottom row includes only particles being member
    of a force chain. From left to right: $\Omega = 2$\,Hz, $\Omega =
    10$\,Hz, $\Omega = 19$\,Hz. By visual inspection it becomes
    obvious that most of the heavier loaded particles belong to force
    chains and are found near to the wall of the mill.}
  \label{volk:dist}
\end{figure}
\end{onecolumn}
\begin{twocolumn}
  
\end{twocolumn}

Surprisingly, for lower rotation velocities one finds that direct
impacts with high relative velocities at the material surface do not
result in high local pressure. Therefore, these impacts are not
relevant for the fragmentation of grains in ball mills. Regions of
high pressure can be found mainly near to the wall, deep inside the
material.  This observation is in good agreement with the experimental
result of Rothkegel and Rolf~\cite{Rothkegel:1992,Rolf}. The extreme right
figure in Fig.~\ref{volk:ldist} reveals a different behavior. The
maximum is near to the material surface where heavy impacts occur. The
absolute values of the pressure are much lower than in the left
figures. This observation relates to the existence of rather weak
force chains.

Figs.~\ref{volk:dist} show the spatial distribution of the number of
particles, which experience a pressure $P_i > 3000$\,g\,cm\,s$^{-2}$.
In these figures the particles which belong to a force chain are only
considered in the bottom row whereas particles not belonging to a
force chain are contained in the top row. Figs.~\ref{volk:dist} can be
directly compared with the figures of Rothkegel~\cite{Rothkegel:1992}
(see Fig.~\ref{volk:ing}); both are found to be in good agreement. The
maxima of the distribution for $\Omega_{\rm I}$ and $\Omega_{\rm II}$
lie near to the wall of the mill, inside the material as observed in
the experiment. Those maxima are an exclusive result of force chains.
Grains not belonging to a force chain contribute only weakly to the
comminution. For the higher velocity $\Omega_{\rm III}$ direct impacts
of particles at the surface are the dominant part. Nevertheless,
absolute values are much lower than for $\Omega_{\rm I}$ and
$\Omega_{\rm II}$. Again, this observation is in good agreement with
the experiment\cite{Rothkegel:1992}.
\section{Conclusion}
We have simulated the process of autogenous 
dry comminution in a ball
mill by the method of molecular dynamics. To this end we developed a
novel algorithm which accounts for fragmentation of particles.
Throughout our simulations all particles, including the fragments,
were modelled as spheres. To accomplish this idealization we
introduced rather short-lived unphysical transient situations in which
the fragments resulting from a cracked particle penetrate each other.
The results of our simulations are available in the internet as
animated sequences under
http://summa.physik.hu-berlin.de/$\sim$kies/mill/bm.html

Employing our molecular dynamics algorithm we could satisfactorily
reproduce experimental phenomena, e.g.~the normalized comminution rate
as a function of the normalized rotation speed and the spatial
pressure distribution.

From the achieved numerical data we were able to explain an
experimental result which was poorly understood so
far~\cite{Rothkegel:1992}: Comminution processes in ball mills occur
deep inside the material rather than close to the surface where the
relative collision velocity is maximal. We explained this effect by
the formation of force chains. Particles which are members of one ore
more force chains experience a significantly larger average force,
thus substantially enhancing the fragmentation probability, as
compared to particles which do not belong to any force chain.

From this we concluded that the experimentally observed and
numerically achieved spatial pressure and fragmentation distributions
are crucially determined by the formation of force chains. Direct
collisions of particles are of minor importance for comminution.
Without the existence of force chains ball mills would work much less
efficient. In the context of a ball mill efficiency optimization it
would be intriguing to investigate the possibility of enhancing
the formation of force chains, e.g.~by modifications of the
geometrical properties of mills, additional application of vibration
etc.

The method developed in this article may be elaborated in many different
aspects:
\begin{itemize}
\item The numerical investigations were done in two dimensions. Ball
  mills are three-dimensional devices.
\item Axial transport of the material was not simulated.
\item All particles, including fragments, were spherical.
\item Only a single-level process was simulated. Modern ball mills
  frequently operate in multi-level mode.
\item Adhesive forces were not considered in the force law.
\item Variation of material properties which originates from
  an increase of temperature due to comminution was not considered.
\end{itemize}

We expect that molecular dynamics will rapidly develop towards an
important  tool for the powder technology machinery. Due
to the rapid progress in hardware and software technology, including
parallel algorithms, it will be possible soon to construct and
optimize powder technology facilities.
\section*{Acknowledgement}
The authors thank S.~Bernotat, R.~Dr\"ogemeier, E.~Gommeren,
G.~Gudehus, H.~J.~Herrmann, L.~Rolf, J.~Schwedes and R.~Weichert for
discussion and for providing relevant literature. The work was
supported by German Science Foundation through project Po472/3-2.

\end{twocolumn}

\begin{thebibliography}{99}
\bibitem{ballmill}
A tumbling ball mill is a tubular vessel which is kept rotating around
its own axis. The grinding medium, here balls, is elevated by the
vessel wall and then rolls, slides, or falls toward the lowest point
of the vessel. The feed material is trapped between the media
particles and, thus, ground by their movement. In case the grinding
media is the feed material itself one speaks of autogeneous grinding.
Even though the autogeneous comminution in a tumbling mill is
considered throughout this paper we, nevertheless, will use the simple
term ball mill.
\bibitem{pressure} The term pressure distribution does not relate to
  the spatially averaged force vector field, but to the absolute value
  of forces acting between particles. Strictly speaking, this quantity
  has the unit of a force, however, to avoid confusion with the
  averaged force vector field we adopt the term pressure from
  engineering literature.
\end{thebibliography}
\end{document}